\documentclass[aps,prx,preprint,superscriptaddress]{revtex4-1}

\usepackage{graphicx,amsmath,amsfonts,mathrsfs,epstopdf}

\begin{document}


\title{Mechanisms and origins of half-metallic ferromagnetism in CrO$_2$}


\author{I. V. Solovyev}
\email{SOLOVYEV.Igor@nims.go.jp}
\affiliation{Computational Materials Science Unit,
National Institute for Materials Science, 1-1 Namiki, Tsukuba,
Ibaraki 305-0044, Japan}
\affiliation{
Department of Theoretical Physics and Applied Mathematics, Ural Federal University,
Mira str. 19, 620002 Ekaterinburg, Russia}
\author{I. V. Kashin}
\author{V. V. Mazurenko}
\affiliation{
Department of Theoretical Physics and Applied Mathematics, Ural Federal University,
Mira str. 19, 620002 Ekaterinburg, Russia}


\date{\today}

\begin{abstract}
Using realistic low-energy electron models,
derived from the first-principles electronic structure
calculations, we investigate behavior of interatomic exchange interactions in
CrO$_2$, which is regarded to be one of the canonical
half-metalic (HF) ferromagnetic. For these purposes we employ
the dynamical mean-field theory (DMFT), based on the exact
diagonalization of the effective Anderson impurity Hamiltonian,
which was further
supplemented with the
theory of infinitesimal spin rotations for the exchange interactions.
In order to elucidate the relative roles played by static and dynamic electron correlations,
we compare the obtained results with several static techniques,
including the unrestricted Hartree-Fock (HF) approximation,
static DMFT (corresponding to the infinite frequency limit for the self-energy), and optimized effective
potential (OEP) method for treating the correlation interactions in the random-phase approximation.
Our results demonstrate that
the origin of the HM ferromagnetism in CrO$_2$ is highly nontrivial.
As far as the interactions in neighboring coordination spheres are concerned, HF and DMFT methods produce
very similar results, due to the partial cancelation of ferromagnetic (FM) double exchange and
antiferromagnetic (AFM) superexchange contributions, which represent two leading terms in the $(\Delta \hat{\Sigma})^{-1}$
expansion for the exchange interactions ($\Delta \hat{\Sigma}$ being the intraatomic exchange splitting between
majority- and minority-spin states).
Both contributions are weaker in
the HF approximation due to, respectively, additional orbital polarization of
the $t_{2g}$ states and
neglect of dynamic
correlations.
The role of higher-order terms in the $(\Delta \hat{\Sigma})^{-1}$ expansion is twofold.
On the one hand, they give rise to additional FM contributions to the neighboring exchange interactions,
which tend to stabilize the FM state.
On the other hand, they produce AFM
long-range interactions, which make the FM state unstable
in the DMFT calculations for the minimal model, consisting of the $t_{2g}$ bands.
Thus, the robust ferromagnetism in the $t_{2g}$ model, which can be easily obtained using static approximations, is
fortuitous and this picture is largely revised at the level of more rigorous DMFT approach.
We argue that the main ingredients, which are missing in the $t_{2g}$ model, are the
direct exchange interactions and the magnetic
polarization of the oxygen $2p$ band. We evaluate these contributions in the
local-spin-density approximation and argue that they play a very important role
in stability of the FM ground state in CrO$_2$.
\end{abstract}


\maketitle

\section{\label{sec:Intro} Introduction}

  CrO$_2$ provides a rare example of metallic ferromagnetism in stoichiometric oxides.
It is widely used in magnetic recording and still
considered as one of the best particulate ever invented for these purposes \cite{SkomskiCoey,Skomski}.
Besides magnetorecording, chromium dioxide has attracted a considerable interest due to its
half-metallic (HM) electronic structure, which was predicted by first-principles
calculations \cite{Schwarz}. The HM electronic structure is such that
the majority-spin electrons are metallic, whereas the minority-spin electrons are semiconducting \cite{deGroot}.
In CrO$_2$, such behavior has been supported by point-contact Andreev reflection measurements \cite{Soulen}.
Because of its implication in various spin-dependent transport phenomena \cite{Singh},
the half-metallicity is the very important property of magnetic substances,
which is intensively studied today \cite{HMRevModPhys}. These studies
typically include both fundamental and practical aspects.

  Needless to say that ferromagnetism is one of the key properties of CrO$_2$, which predetermines
its popularity and importance in all the applications. Although the Curie temperature is not exceptionally high
from the view point of practical applications (about $390$ K, meaning that the magnetic properties
are significantly deteriorated at room temperature) \cite{Skomski}, it is still sufficiently high
in order to classify CrO$_2$ as ``robust ferromagnet''.

  Because of its popularity, CrO$_2$ is the well studied material, both theoretically and experimentally.
There is a fair number of theoretical works, focusing on the analysis of structural, transport,
optical, and electronic properties of CrO$_2$ \cite{Schwarz,Sorantin,Lewis,Korotin,Mazin,Yamasaki,Chioncel,Craco}.
Many of them are based on the first-principles electronic
structure calculations. These works clarify many important aspects of the
material properties of CrO$_2$. However, despite its immanent importance in the field,
the problem of interatomic magnetic interactions and stability of the ferromagnetic (FM) ground state is CrO$_2$ remains
in the shadow. Particularly, why is CrO$_2$ ferromagnetic? What are the main microscopic mechanisms yielding the
FM ground state in CrO$_2$? From our point of view, these important questions remain largely unanswered
and in the present work we will try to fill in this gap.

  The ferromagnetism in CrO$_2$ is typically ascribed to the double exchange (DE) mechanism \cite{Korotin,Schlottmann,Laad},
which was originally introduced for magnetoresistive manganites \cite{Zener,AndersonHasegawa,deGennes,Dagotto}.
This mechanism is governed by the large
intraatomic exchange splitting ($\Delta \hat{\Sigma}$) between the majority ($\uparrow$) and minority
($\downarrow$) spin states, which penalizes the electron hoppings
between atoms with opposite directions of spins. In such situation,
the FM state will be the most stable one because
any deviation from
the collinear FM alignment of spins will increase the kinetic energy of electrons.
The DE picture is well justified for large-spin ($S$) systems.
In manganites, where $S=2$, it can be very useful for semi-quantitative analysis, and, in many cases, provides
a valuable insight in understanding the electronic and magnetic properties \cite{Dagotto,Springer}. However,
even in this case, the additional effects can play a very
important role and substantially modify the canonical DE picture \cite{PRL99}. For instance, the well-known
antiferromagnetic (AFM) superexchange interaction \cite{PWA}, which is also important in manganites,
is formally a next-order effect
in the $(\Delta \hat{\Sigma})^{-1}$ expansion for interatomic exchange interactions \cite{Springer,PRL99}. In CrO$_2$, where $S=1$,
the interatomic exchange splitting is not particularly large and the DE picture can be
even more problematic: namely, besides FM DE interactions between the nearest neighbors, one can expect other
magnetic interactions (not necessarily the FM ones), which can alter the magnetic ground state \cite{Springer}.
Another important factor, which is not treated by the DE model, is the oxygen states \cite{Oguchi,Priya,Ku,Mazurenko}.

  Another disputable point is the role of electron correlations beyond the local-spin-density approximation (LSDA) and
whether CrO$_2$ should be regarded as a strongly-correlated material or not. On the one hand, LSDA and generalized gradient approximation (GGA)
already provide a reasonable description for the structural, transport,
and optical properties of CrO$_2$ with only moderate manifestation of many-body effects \cite{Sorantin,Lewis,Mazin}.
On the other hand, it was also suggested that electron correlations are essential for understanding
results of photoemission, x-ray absorption, optical, and resistivity measurements \cite{Craco}.
We are not aware of any investigation of the effect of electron correlations on the behavior of
interatomic exchange interactions in CrO$_2$. Basically, there is only one theoretical work \cite{Sims},
which addresses the problem of interatomic exchange interactions in CrO$_2$ on the basis of first-principles GGA and LSDA$+$$U$
calculations. However, both are static techniques and do not treat dynamic correlations.
Moreover, the reliability of the LSDA$+$$U$ approach suffers from the use of adjustable parameters
and still unresolved problem of how to construct the LSDA$+$$U$ functional in order to
avoid the
double-counting problem \cite{PRB98}.
Taking into account the above controversy, it is crucially important to treat the electron correlations (if any) in the most
unambiguous manner. In the present work, we will try to pursue this strategy, first, by constructing the realistic model for CrO$_2$
and deriving all the parameters from first-principles calculations and, then, by solving this model within the
dynamical mean-field theory (DMFT), supplemented with the exact diagonalization (ED) method for the quantum impurity problem.
We will show that stability of the HM FM ground state in CrO$_2$ is a highly nontrivial problem. If one considers only
static electron correlations in the frameworks of either unrestricted Hartree-Fock (HF) or static DMFT techniques, the
FM ground state can be formally obtained already in the minimal model, including only the closest to the Fermi level $t_{2g}$ bands.
However, the dynamic correlations tend to destabilize this state. Therefore, in order to explain the experimentally observed
ferromagnetism in CrO$_2$, it is crucially important to consider other magnetic interactions and we argue that these are the direct exchange
interactions between Wannier functions centered at different Cr sites and the polarization of the oxygen $2p$ band.
Another static approach -- the so-called optimized effective
potential (OEP) method, treating the correlation interactions in the random-phase approximation, produces a curious but unphysical
insulating solution and further suppresses the tendencies towards ferromagnetism. This again emphasizes the importance of consistent treatment of
correlation interactions in CrO$_2$.

  The rest of the article is organized as follows. In Sec.~\ref{sec:Method} we will explain details of our method:
the construction of effective low-energy model (Sec.~\ref{sec:Method}), the solution of DMFT equations (Sec.~\ref{sec:DMFT}),
and the difference between unrestricted HF and static DMFT techniques. In Sec.~\ref{sec:J} we will present our results for
interatomic exchange interactions and discuss them in many details: the $(\Delta \hat{\Sigma})^{-1}$ expansion
for
nearest-neighbor (NN) and next-NN interactions (Sec.~\ref{sec:DE}),
the magnetic state dependence of interatomic exchange interactions (Sec.~\ref{sec:finite}),
the behavior of long-range interactions (Sec.~\ref{sec:JLR}),
the contribution of the direct exchange interactions and the oxygen states (Sec.~\ref{sec:Oband}),
as well as results of the OEP method (Sec.~\ref{sec:dcorr}). Finally, in Sec.~\ref{sec:Summary}, we will present a
summary of our work.

\section{\label{sec:Method} Method}

\subsection{\label{sec:LEmodel} Parameters of effective low-energy model}

  In this section, we briefly remind the reader
the main ideas behind the construction of effective low-energy model and
present results of such construction for CrO$_2$.
The methodological details can be found in the review article \cite{review2008}.
All calculations have been performed using parameters of experimental rutile structure
(the space group $P4_2/mnm = D_{4h}^{14}$) \cite{Porta}.

\begin{figure}[h!]
\begin{center}
\includegraphics[width=10cm]{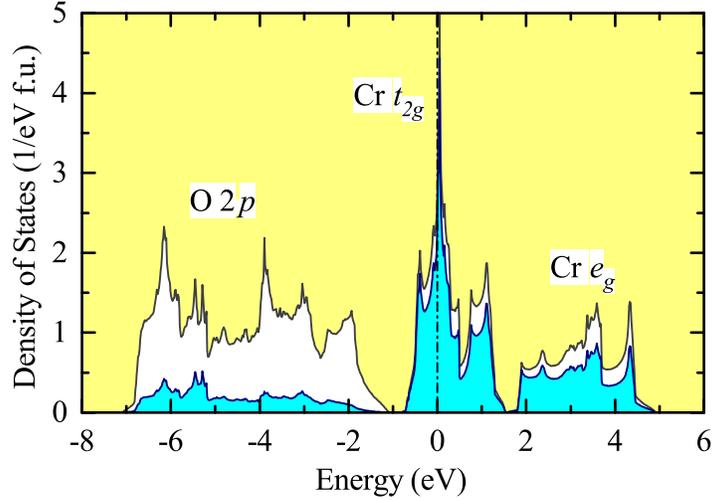}
\end{center}
\caption{\label{fig.DOS}
(Color online) Total and partial densities of states of CrO$_2$ in the local density approximation.
The shaded light (blue) area shows the contribution of the Cr $3d$ states.
The positions of the main bands are indicated by symbols. The Fermi level is at zero energy (shown by dot-dashed line).
}
\end{figure}

\begin{figure}[h!]
\begin{center}
\includegraphics[width=15cm]{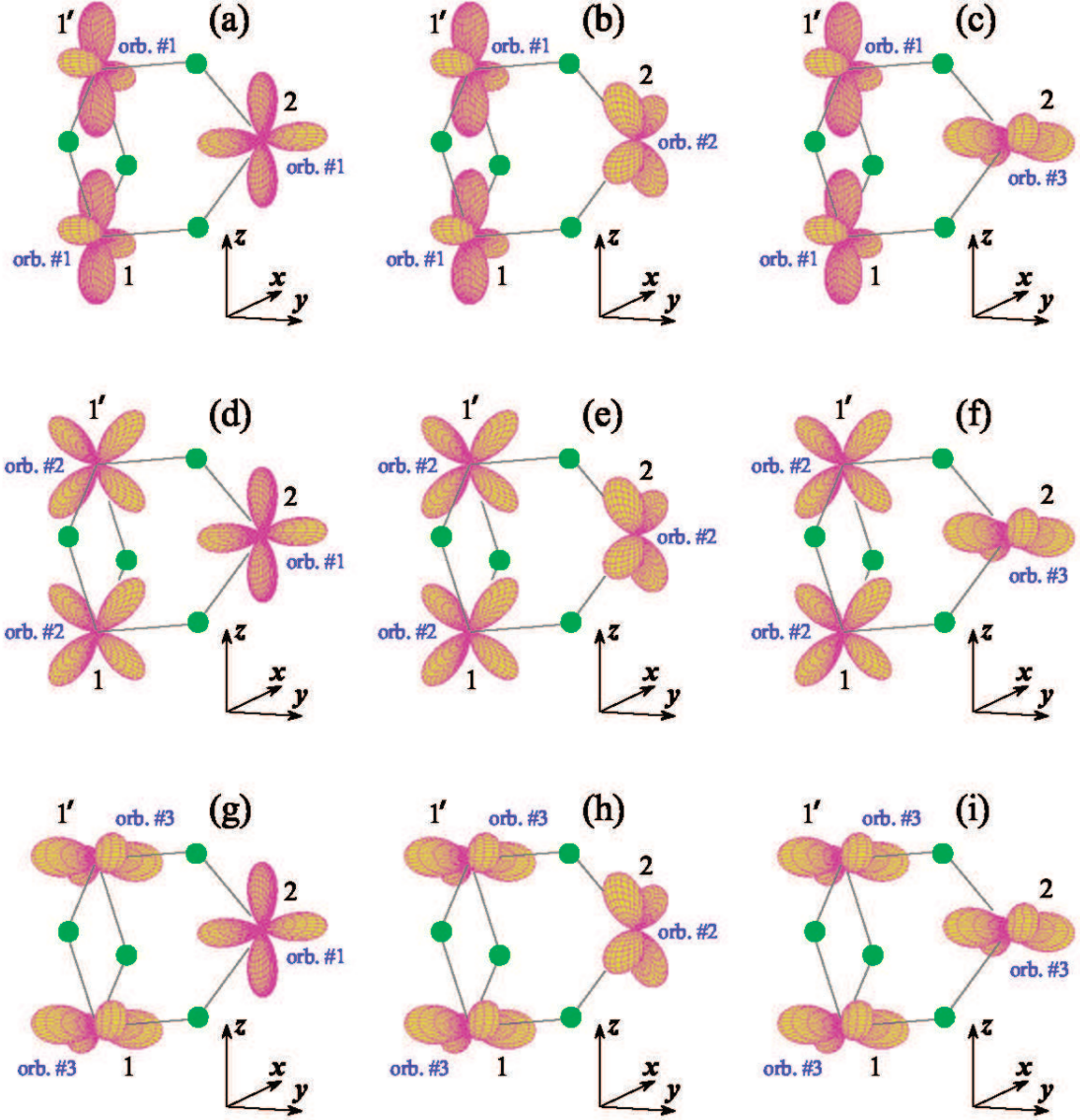}
\end{center}
\caption{\label{fig.orbitals}
(Color online)
Atomic electron densities, explaining
relative positions of Cr $t_{2g}$ orbitals at the sites $1$, $1'$, and $2$.
The oxygen atoms are indicated by the green spheres.
}
\end{figure}

  The model Hamiltonian,
\begin{equation}
\hat{\cal{H}}  =  \sum_{ij} \sum_\sigma \sum_{ab}
t_{ij}^{ab}\hat{c}^\dagger_{i a \sigma}
\hat{c}^{\phantom{\dagger}}_{j b \sigma} +
  \frac{1}{2}
\sum_{i}  \sum_{\sigma \sigma'} \sum_{abcd} U^i_{abcd}
\hat{c}^\dagger_{i a \sigma} \hat{c}^\dagger_{i c \sigma'}
\hat{c}^{\phantom{\dagger}}_{i b \sigma}
\hat{c}^{\phantom{\dagger}}_{i d \sigma'},
\label{eqn.ManyBodyH}
\end{equation}
is formulated
in the basis of Wannier orbitals $\{ \phi_{i a} \}$,
which are constructed for the magnetically active Cr $t_{2g}$ bands near the Fermi level,
starting from the band structure in the local-density approximation (LSDA) (Fig.~\ref{fig.DOS}).
Here,
$\sigma (\sigma')$$=$ $\uparrow$/$\downarrow$ are the spin indices, while $a$, $b$, $c$, and $d$ label
three $t_{2g}$ orbitals, which
have the following form in the global coordinate frame:
$|1 \rangle = \pm$$\frac{1}{2}|xy \rangle$$+$$\frac{\sqrt{3}}{2}|3z^2$$-$$r^2 \rangle$,
$|2 \rangle = \frac{1}{\sqrt{2}}|yz \rangle$$\pm$$|zx \rangle$, and
$|3 \rangle = |x^2$$-$$y^2 \rangle$,
where the upper and lower signs stand for the Cr site 1 and 2, respectively (see Fig.~\ref{fig.orbitals}).
These orbitals are sometimes denoted as, respectively, $| xy \rangle$, $| yz$$-$$zx \rangle$, and
$| yz$$+$$zx \rangle$, referring to the local coordinate frame \cite{Yamasaki}.
It is important that at Cr sites all three orbitals belong to different irreducible representations of the point group $mmm = D_{2h}$,
meaning that all local quantities, including the crystal field, DMFT self-energy, and local Green's function,
will be diagonal with respect to these orbital indices. Moreover, the diagonal matrix elements will be the same
for the Cr-sites $1$ and $2$.

  Each lattice point $i$ ($j$) is specified by the position
$\boldsymbol{\tau}$ ($\boldsymbol{\tau}'$)
of the Cr site in the primitive cell and the lattice translation ${\bf R}$.
Hence, the basis orbital
$\phi_{i a}({\bf r}) \equiv \phi_{\tau a} ({\bf r}$$-$${\bf R}$$-$$\boldsymbol{\tau})$
is centered in
the lattice point $({\bf R}$$+$$\boldsymbol{\tau})$
and labeled by the indices $\tau$ and $a$.
The Wannier functions were calculated using the
projector-operator technique \cite{review2008,WannierRevModPhys}
and orthonormal linear muffin-tin orbitals (LMTO's)  \cite{LMTO1,LMTO2,LMTO3} as the trial wave functions.
In physical terms, LMTO can be viewed as the localized atomic-like Wannier function constructed for the
whole region of valence states.
Therefore, the projector-operator technique allows us to generate
well localized Wannier functions for the $t_{2g}$ bands, that is guaranteed by the good localization of LMTO's themselves.
Then, the one-electron part of the model
is identified with the matrix elements of LDA Hamiltonian (${\cal H}_{\rm LDA}$) in the Wannier basis:
$t^{ab}_{\boldsymbol{\tau}, \boldsymbol{\tau}'+{\bf R}} =
\langle \phi_{\tau a} ({\bf r}$$-$$\boldsymbol{\tau})| {\cal \hat H}_{\rm LDA} |
\phi_{\tau' b} ({\bf r}$$-$${\bf R}$$-$$\boldsymbol{\tau}') \rangle$.
Since the Wannier basis is complete in the low-energy part of the spectrum, the construction is exact in the sense that
the band structure, obtained from $t^{ab}_{\boldsymbol{\tau}, \boldsymbol{\tau}'+\bf R}$,
exactly coincides with the one of LDA.

  The site-diagonal part of $\hat{t}_{ij} \equiv [t_{ij}^{ab}]$ describes the crystal field splitting.
It has the following form (in meV):
\begin{equation}
\hat{t}_{11} =
\left(
\begin{array}{ccc}
 -246 &  0 &   0 \\
    0 & 60 &   0 \\
    0 &  0 & 186 \\
\end{array}
\right).
\label{eqn:t11}
\end{equation}
The matrices of transfer integrals in the bonds $1$-$1'$ and $1$-$2$ are given by
\begin{equation}
\hat{t}_{11'} =
\left(
\begin{array}{ccc}
 -67 &    0 &    0 \\
   0 & -191 &    0 \\
   0 &    0 &  158 \\
\end{array}
\right)
\label{eqn:t11p}
\end{equation}
and
\begin{equation}
\hat{t}_{12} =
\left(
\begin{array}{ccc}
 -15 &    0 &    0 \\
 -28 &    0 &    0 \\
   0 &  194 & -119 \\
\end{array}
\right),
\label{eqn:t12}
\end{equation}
respectively. Other transfer integrals are considerably weaker \cite{note1}.
The obtained values are
in reasonable agreement with results of previous calculations \cite{Yamasaki}.
One interesting aspect
is the large matrix element $t_{11'}^{33} = 158$ meV, which is formally
of the $dd\delta$ type (see Fig.~\ref{fig.orbitals}) and, therefore, supposed to be weak \cite{SlaterKoster}.
Nevertheless, such large transfer integrals are possible due to peculiar geometry of the CrO$_2$ lattice
and contributions of intermediate O $2p$ and Cr $e_g$ states \cite{Yamasaki}.
In terms of the Wannier functions, this means that the functions should have a sizable tail spreading to the oxygen and other
Cr sites \cite{Yamasaki}. Thus, already from this fact one can expect appreciable direct exchange interactions, which will
be evaluated in Sec.~\ref{sec:Oband}.
Another interesting aspect is the large asymmetric contribution $t_{12}^{32} = 194$ meV, caused by the
electron transfer via intermediate oxygen atom (see Fig.~\ref{fig.orbitals}h).
The same mechanism is responsible for finite $t_{12}^{21}$. However, it is considerably smaller than $t_{12}^{32}$.

  Matrix elements of the on-site Coulomb interactions can be also obtained in the Wannier basis as
$$
U_{abcd} = \int d {\bf r} \int d {\bf r}'
\phi_{i a}^* ({\bf r})
\phi_{i b} ({\bf r}) v_{\rm scr}({\bf r},{\bf r}')
\phi_{i c}^* ({\bf r}') \phi_{i d} ({\bf r}'),
$$
where the screened interaction
$v_{\rm scr}({\bf r},{\bf r}')$ is computed in the constrained
random-phase approximation (RPA) \cite{Ferdi04}.
Since RPA is very time consuming, we apply additional approximations, which were discussed
in \cite{review2008}. Namely, first we evaluate the screened Coulomb and exchange interactions between
atomic Cr $3d$ orbitals, using fast and more suitable for these purposes constrained LDA technique. After that,
we consider additional channel of screening caused by the $3d \rightarrow 3d$ transitions
in the polarization function of constrained RPA and project this
function onto the $3d$ orbitals. The so obtained parameters of Coulomb interactions
are well consistent with results of full-scale constrained RPA calculations
without additional approximations.

  The obtained matrices of the on-site Coulomb interactions were fitted in terms of two
Kanamori parameters \cite{Kanamori}:
the parameter of intra-orbital Coulomb interaction ${\cal U} = 2.84$ eV and the exchange interaction ${\cal J} = 0.70$ eV.
The third Kanamori parameter -- the so-called inter-orbital Coulomb interaction -- can be obtained from ${\cal U}$ and ${\cal J}$ as
${\cal U}' = {\cal U} - 2{\cal J}$.

\subsection{\label{sec:DMFT} Dynamical mean-field theory}

  Solution of the low-energy model, represented by Hamiltonian (\ref{eqn.ManyBodyH}), is a complicated numerical and methodological problem.
In general, microscopic properties of a periodic magnetically collinear system can be expressed via
one-electron Green's function $\hat{G}^{\uparrow, \downarrow}(\omega, {\bf k})$,
which, in the reciprocal space, can be formally related to the frequency- and momentum-dependent
self-energy $\hat{\Sigma}^{\uparrow, \downarrow}(\omega, {\bf k})$:
\begin{eqnarray}
\hat{G}^{\uparrow, \downarrow}(\omega, {\bf k}) =
\left[\omega - \hat{t}({\bf {k}}) - \hat{\Sigma}^{\uparrow, \downarrow}(\omega, {\bf k}) \right]^{-1},
\label{eqn.latticeG}
\end{eqnarray}
where ${\hat{t}}({\bf {k}})$ is the one-electron part of the Hamiltonian (\ref{eqn.ManyBodyH})
in the reciprocal space and all kind of correlation effects are described by $\hat{\Sigma}^{\uparrow, \downarrow}(\omega, {\bf k})$.

  The basic approximation, underlying the dynamical mean-field theory (DMFT),
is that the self-energy is assumed to be independent on ${\bf k}$:
\begin{eqnarray}
\hat{\Sigma}^{\uparrow, \downarrow}(\omega, {\bf k}) \approx \hat{\Sigma}^{\uparrow, \downarrow}(\omega),
\end{eqnarray}
which becomes exact in the limit of infinite dimensions (or coordination numbers) \cite{DMFTRevModPhys}.
The main idea of DMFT is to map the initial many-body problem for the crystalline lattice onto the quantum impurity one,
surrounded by an effective electronic bath, and find self-consistently the parameters of this bath.
Namely, the local (or site-diagonal) Green function of the crystal is given by
$$
\hat{G}^{\uparrow, \downarrow}(\omega) = \sum_{{\bf k}} \hat{G}^{\uparrow, \downarrow}(\omega, {\bf k}).
$$
It can be further used to obtain the bath Green function, ${\cal G}(\omega)$, from the Dyson equation:
\begin{eqnarray}
\hat{G}^{\uparrow, \downarrow}(\omega) =
\hat{\cal G}(\omega) + \hat{\cal G}(\omega) \hat{\Sigma}^{\uparrow, \downarrow}(\omega) \hat{G}^{\uparrow, \downarrow}(\omega).
\label{eqn.Dyson}
\end{eqnarray}

  Then, new $\hat{G}^{\uparrow, \downarrow}(\omega)$ is obtained by solving the Anderson impurity model. The
corresponding Hamiltonian is given by
\begin{eqnarray}
\hat{\cal{H}}_{imp} = \sum_{a \sigma} E_a \hat{d}^{\dagger}_{a \sigma} \hat{d}_{a \sigma}
+ \frac{1}{2}
\sum_{abcd, \sigma, \sigma'} U_{abcd} \hat{d}^{\dagger}_{a \sigma} \hat{d}^{\dagger}_{c \sigma'} \hat{d}_{b \sigma} \hat{d}_{d \sigma'}
\label{eqn.Anderson} \\
+ \sum_{ap,\sigma} \left[ V_{ap} \hat{d}_{a \sigma}^{\dagger} \hat{c}_{p \sigma} + H.c. \right]
+ \sum_{p,\sigma} \epsilon_p \hat{c}_{p \sigma}^{\dagger} \hat{c}_{p \sigma}, \nonumber
\end{eqnarray}
where $\hat{d}(\hat{d}^{\dagger})$
and $\hat{c}(\hat{c}^{\dagger})$ are the electron annihilation(creation) operators for the impurity and bath states, respectively,
$V_{ap}$ is impurity-bath hybridization, and $E_a$ ($\epsilon_p$) are the noninteracting energy levels of the impurity (bath).
In order to obtain parameters of the Anderson impurity model, we adapt the following analytical form of the
bath Green function (separately for each $t_{2g}$ orbital $a$):
$$
{\cal {G}}_{a}^{N_{s}}(\omega) = \left( \omega - E_a - \sum_{p} \frac{V_{ap}^{2}}{\omega - \epsilon_p} \right)^{-1}
$$
and fit it in terms of $E_a$, $\epsilon_p$, and $V_{ap}$. Generally speaking, the number of bath states $p$ is infinite.
However, in order to handle this problem numerically by means of ED, we discretize the bath and use a finite number of
electronic orbitals $N_{s}$. It enables us to numerically diagonalize the impurity Hamiltonian (\ref{eqn.Anderson}) and
obtain $\hat{\cal G}^{\uparrow, \downarrow}_{imp}$, which is further identified with $\hat{G}^{\uparrow, \downarrow}(\omega)$.
Then, using ${\cal {G}}_{a}(\omega) \equiv {\cal {G}}_{a}^{N_{s}}(\omega)$, the new self-energy can be found from
the Dyson equation (\ref{eqn.Dyson}). After that, it is substituted into Eq.~(\ref{eqn.latticeG}) to obtain new
$\hat{G}^{\uparrow, \downarrow}(\omega, {\bf k})$, and the problem is solved self-consistently.

   The ED method allows us to find the ground state as well as the low-lying excitations of the quantum impurity model.
The standard numerical algorithms to treat the eigenproblem are based on the matrix-vector multiplication,
where the initial vector, matrix and net vector are kept in computer's random-access memory (RAM).
In order to make our model treatment realistic, it is necessary to take total $N_s$ ranging from 15 to 18.
However, it would lead to
the Hamiltonian matrix (\ref{eqn.Anderson}) of the dimensionality $\sim(10^{10}$$\times$$10^{10})$,
which makes the diagonalization procedure troublesome, even for modern multiprocessor computers.

\begin{figure}[h!]
\begin{center}
\includegraphics[width=12cm]{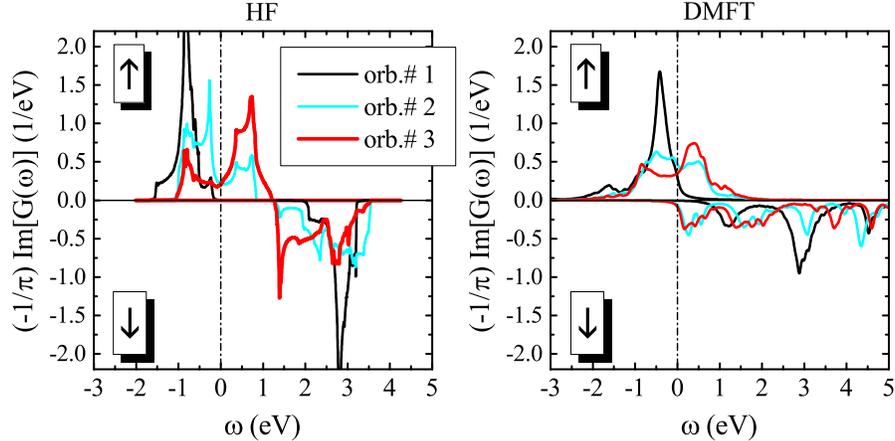}
\end{center}
\caption{\label{fig.DOSmodel}
(Color online) Partial densities of states as obtained in the unrestricted
Hartree-Fock approach (left) and the dynamical mean-field theory (right)
for the ferromagnetic state.
The Fermi level is at zero energy (shown by dot-dashed line).
}
\end{figure}

  In this study we use the newly developed numerical ED scheme,
based on the standard Arnoldi algorithm implemented in ARPACK program package \cite{ARPACK},
where the Hamiltonian matrix is not stored in the RAM, but efficiently
recalculated `on the fly' at each matrix-vector multiplication step.
It makes the computational time to increase by only 10-15\%.
However, the amount of necessary RAM is decreased by 80-90\%,
giving us the possibility to perform realistic calculations with
the large number of effective orbitals $N_{s}$. Particularly,
using this scheme, we were able to include 4 bath states per each $t_{2g}$ orbital in the framework of DMFT.
We have confirmed that the obtained electronic structure is well converged depending on the
number of the bath states. The numerical calculations have been performed for the temperature
$T = 232$ K, which is substantially larger than the magnetic transition temperature,
and in the external magnetic field $\mu_{\rm B}H = 5$  meV, which is required in order to lift the
magnetic degeneracy of multiplet states \cite{Mazurenko1}.
The example of electronic spectrum is shown in Fig.~\ref{fig.DOSmodel},
which is in remarkable agreement with results of the previous DMFT studies \cite{Chioncel}.

\subsection{\label{sec:SDMFT} Static DMFT versus unrestricted Hartree-Fock approach}

  The asymptotic high-frequency behavior of $\hat{\Sigma}^{\uparrow,\downarrow}(\omega)$
in DMFT is given by \cite{WangDangMillis}:
\begin{equation}
\Sigma_1^{\uparrow}(\infty) = ({\cal U}-3{\cal J})(n_2^{\uparrow} + n_3^{\uparrow})
+ {\cal U}n_1^{\downarrow}
+ ({\cal U}-2{\cal J})(n_2^{\downarrow} + n_3^{\downarrow}),
\label{eqn:Sinfinity}
\end{equation}
where $\{ n_a^{\uparrow,\downarrow} \}$ are the self-consistent populations in DMFT:
$$
n_a^{\uparrow, \downarrow} = - \frac{1}{\pi} {\rm Im}
\int_{- \infty}^{\varepsilon_{\rm F}} d \omega \, G^{\uparrow, \downarrow}_a (\omega).
$$
Other matrix elements of $\hat{\Sigma}^{\uparrow,\downarrow}(\infty)$
can be obtained from Eq.~(\ref{eqn:Sinfinity}) by permutation of the spin and orbital indices.
$\hat{\Sigma}^{\uparrow,\downarrow}(\infty)$ has the same form as the
potential matrix in the unrestricted Hartree-Fock method \cite{review2008}, but with different populations:
in DMFT, these populations include the effect of frequency-dependence of the self-energy, while in HF, they do not.
For the FM state, these populations are summarized in Table~\ref{tab:n}.
\begin{table}[h!]
\caption{Self-consistent orbital populations $\{ n_a^{\uparrow,\downarrow} \}$ for the ferromagnetic states,
as obtained in DMFT and unrestricted Hartree-Fock calculations for the low-energy model.}
\label{tab:n}
\begin{ruledtabular}
\begin{tabular}{ccccccc}
 & $n_1^{\uparrow}$ & $n_2^{\uparrow}$ & $n_3^{\uparrow}$ &  $n_1^{\downarrow}$ & $n_2^{\downarrow}$ & $n_3^{\downarrow}$ \\
\hline
 DMFT & $0.934$ & $0.587$ & $0.431$ & $0.010$ & $0.020$ & $0.023$ \\
 HF   & $0.999$ & $0.710$ & $0.291$ & $0$ & $0$ & $0$ \\
\end{tabular}
\end{ruledtabular}
\end{table}
Besides small population of the $\downarrow$-spin states (and, therefore, small deviation from the HM behavior),
the main difference between DMFT and HF is in the orbital polarization of the $\uparrow$-spin states
(see Fig.~\ref{fig.DOSmodel}).
The first
orbital is practically fully occupied in both approaches. The population of other two orbitals tends to be nearly equal in DMFT,
while in HF these states are strongly polarized and there is an additional
redistribution of electrons between $n_2^{\uparrow}$ to $n_3^{\uparrow}$.
Finite values of $\{ n_a^{\downarrow} \}$ is the natural result of DMFT calculations for the
HM magnets, which is related to the existence of nonquasiparticle $\downarrow$-spin
states near the Fermi level \cite{HMRevModPhys,Chioncel}.
Nevertheless, we have found that these states have a minor effect on the behavior of
interatomic exchange interactions and, from this point of view, CrO$_2$ can be treated as HM
ferromagnet, even in DMFT.

\section{\label{sec:J} Interatomic exchange interactions}

  We consider the mapping of electron model (\ref{eqn.ManyBodyH}) onto Heisenberg model
with $S=1$:
$$
\hat{\cal{H}}_S = -\frac{1}{2} \sum_{ij} J_i \hat{\boldsymbol{S}}_{j} \cdot \hat{\boldsymbol{S}}_{j+i}.
$$
In these notations, $J_i$ is the exchange coupling between two Cr sites, located in the origin ($0$)
and in the point $i$ of the lattice, relative to the origin.
The mapping onto the spin model implies the adiabatic motion of spins when all instantaneous changes of the
electronic structure adjust slow rotations of the spin magnetic moments.
The parameters of this model can be obtained by
using the theory of infinitesimal spin rotations \cite{JHeisenberg,Katsnelson2000}:
\begin{equation}
J_i = \frac{1}{2\pi} {\rm Im} \int_{- \infty}^{\varepsilon_{\rm F}} d \omega \, {\rm Tr}_L \left\{
\Delta \hat{\Sigma}(\omega) \hat{G}_{0i}^{\uparrow}(\omega)
\Delta \hat{\Sigma}(\omega) \hat{G}_{i0}^{\downarrow}(\omega)
\right\},
\label{eqn:Jij}
\end{equation}
where $\hat{G}^{\uparrow, \downarrow}_{0i}(\omega) = [ \omega - \hat{t} -
\hat{\Sigma}^{\uparrow, \downarrow}(\omega) ]^{-1}_{0i}$
is the one-electron Green function between sites $0$ and $i$,
$\Delta \hat{\Sigma} = \hat{\Sigma}^\uparrow - \hat{\Sigma}^\downarrow$ and ${\rm Tr}_L$ denotes the trace over orbital indices.
The parameters $\{ J_i \}$ given by Eq.~(\ref{eqn:Jij}) are nothing but the second derivatives
of the total energy with respect to the rotations of spins. Therefore, this definition of the Heisenberg model is valid only for small rotations of
the magnetic moments near the FM state and characterizes the local stability of this state.
The effect of finite rotations will be discussed in Sec.~\ref{sec:finite}.

\begin{figure}[h!]
\begin{center}
\includegraphics[width=12cm]{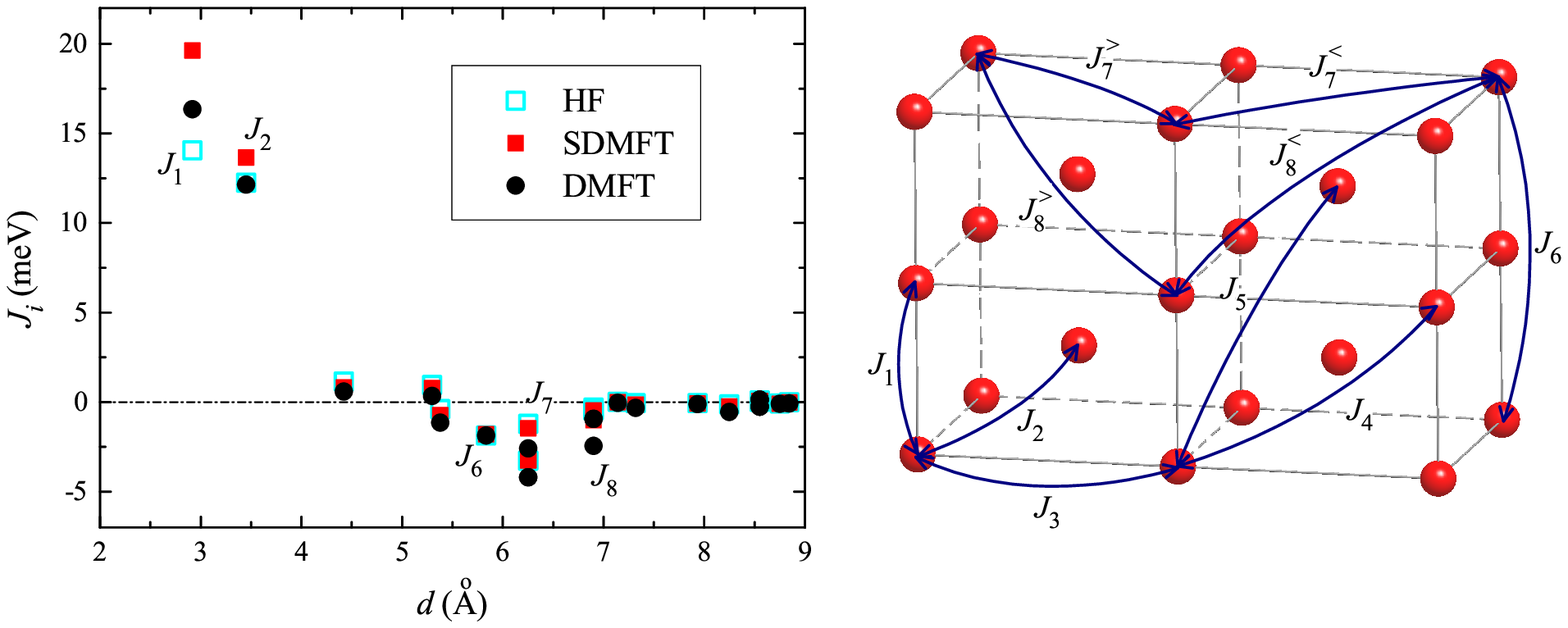}
\end{center}
\caption{\label{fig.J}
(Color online) (Left) Distance dependence of interatomic exchange interactions
as obtained in the theory of infinitesimal spin
rotations with three different types of approximations for the self energy:
the unrestricted Hartree-Fock approximation (HF), the dynamical mean-field theory (DMFT), and the
static limit for the DMFT self-energy $\hat{\Sigma}(\omega$$\to$$\infty)$ (SDMFT).
(Right) Lattice of Cr sites with the notation of interatomic exchange interactions.
}
\end{figure}

  The parameters of interatomic magnetic interactions, obtained in the theory of infinitesimal spin rotations,
are listed in Table~\ref{tab:J} and their behavior is explained in Fig.~\ref{fig.J}.
\begin{table}[h!]
\caption{Parameters of interatomic exchange interactions (in meV)
as obtained in the theory of infinitesimal spin
rotations with three different types of approximations for the self-energy:
the unrestricted Hartree-Fock approximation (HF), the dynamical mean-field theory (DMFT), and the
static limit for the DMFT self-energy $\hat{\Sigma}(\omega$$\to$$\infty)$ (SDMFT).
The notations of parameters $J_i$ are explained in Fig.~\ref{fig.J}.}
\label{tab:J}
\begin{ruledtabular}
\begin{tabular}{crrr}
 parameter  & HF~     &   DMFT  & SDMFT \\
\hline
  $J_1$     & $14.06$ & $16.35$~ & $19.63$~ \\
  $J_2$     & $12.26$ & $12.14$~ & $13.65$~ \\
  $J_3$     &  $1.16$ &  $0.60$~ &  $0.82$~ \\
  $J_4$     &  $0.96$ &  $0.35$~ &  $0.78$~ \\
  $J_5$     & $-0.39$ & $-1.15$~ & $-0.72$~ \\
  $J_6$     & $-1.87$ & $-1.85$~ & $-1.78$~ \\
  $J_7^<$   & $-1.21$ & $-2.58$~ & $-1.45$~ \\
  $J_7^>$   & $-3.26$ & $-4.19$~ & $-3.25$~ \\
  $J_8^<$   & $-0.31$ & $-0.94$~ & $-0.47$~ \\
  $J_8^>$   & $-0.46$ & $-2.44$~ & $-1.00$~ \\
\end{tabular}
\end{ruledtabular}
\end{table}
We note the following: (i) As expected, the FM ground state is stabilized by the NN
and next-NN interactions ($J_1$ and $J_2$, respectively). The values of these interactions,
obtained in DMFT and unrestricted HF approach, are surprisingly close, while static DMFT overestimates
both of them; (ii) Besides strong FM interactions $J_1$ and $J_2$, there are several types
of AFM interactions, operating in the 5th, 6th, 7th, and 8th coordination spheres,
which tend to destabilize the FM state. These interactions are especially strong in the case of DMFT.
In the next sections, we will elucidate the microscopic origin of such behavior and its consequences
on the properties of CrO$_2$.

\subsection{\label{sec:DE} Double exchange and beyond}

  The ferromagnetism of CrO$_2$ is frequently attributed to the DE mechanism \cite{Korotin,Schlottmann,Laad}.
This is the very important point, which needs to be clarified.

  In the HM regime,
all poles of $\hat{G}^{\downarrow}(\omega)$ are located in the unoccupied part of the
spectrum and
$\Delta \hat{\Sigma}$ can be regarded as a large parameter.
Note that the existence of the small weight of nonquasiparticle $\downarrow$-spin states near the Fermi level
does not alter this conclusion, which will remain true even in the case of DMFT.
This justifies the use of the $(\Delta \hat{\Sigma})^{-1}$ expansion in the occupied part \cite{PRL99}:
$$
\hat{G}^{\downarrow}(\omega) = - \Delta \hat{\Sigma}^{-1} \sum_{n=0}^{\infty}
\left( [\Delta \hat{\Sigma} \hat{G}^{\uparrow}]^{-1} \right)^n,
$$
which follows from the identity
$\hat{G}^{\downarrow} = \left( [\hat{G}^{\uparrow}]^{-1} - \Delta \hat{\Sigma} \right)^{-1}$.
The $n=0$ term of this expansion contains only site-diagonal elements and, consequently,
does not contribute to Eq.~(\ref{eqn:Jij}). Therefore, in the HM state, $J_i$ can be presented
as an infinite series:
\begin{equation}
J_i = \sum_{n=1}^{\infty} J_i^{(n)},
\label{eqn:nseries}
\end{equation}
where
\begin{equation}
J_i^{(n)} = - \frac{1}{2\pi} {\rm Im} \int_{- \infty}^{\varepsilon_{\rm F}} d \omega \, {\rm Tr}_L \left\{
\hat{G}_{0i}^{\uparrow}(\omega) \left( [\Delta \hat{\Sigma}(\omega) \hat{G}^{\uparrow}(\omega)]^{-1} \right)^n_{i0} \Delta \hat{\Sigma}(\omega)
\right\}.
\label{eqn:nterm}
\end{equation}
The $n=1$ term corresponds to the DE interaction,
which can be easily found analytically \cite{PRL99}:
\begin{equation}
J_i^{(1)} = \frac{1}{2\pi} {\rm Im} \int_{- \infty}^{\varepsilon_{\rm F}} d \omega \, {\rm Tr}_L \left\{
\hat{G}_{0i}^{\uparrow}(\omega) \hat{t}_{i0}
\right\}.
\label{eqn:JDE}
\end{equation}
Moreover, using the identity $\hat{G}^{\uparrow}(\omega)\left[\omega - \hat{t} -
\hat{\Sigma}^{\uparrow}(\omega) \right] = \hat{1}$,
it is straightforward to show that
\begin{equation}
\sum_i J_{i}^{(1)} = -\frac{1}{2} E_{\rm kin},
\label{eqn:DEEkin}
\end{equation}
where
$E_{\rm kin}$ is the kinetic energy (per one Cr site):
$$
E_{\rm kin} =
- \frac{1}{\pi} {\rm Im} \int_{- \infty}^{\varepsilon_{\rm F}} d \omega \, {\rm Tr}_L \left\{
\hat{G}_{00}^{\uparrow}(\omega) \left[ \omega - \hat{t}_{00} - \hat{\Sigma}^{\uparrow}(\omega) \right]
\right\}.
$$

  From Eq.~(\ref{eqn:DEEkin}) it is clear that the main interactions $J_{i}^{(1)}$ should be positive (or ferromagnetic).
This equation is nearly perfectly reproduced by our calculations.
For instance, we have obtained the following values of DE interactions in DMFT:
$J_1^{(1)} = 28.95$ meV, $J_2^{(1)} = 19.44$ meV,
$J_3^{(1)} = 1.33$ meV, and $J_4^{(1)} = 1.59$ meV. Since the Wannier functions are localized
and the transfer integrals connecting more remote sites are small, the corresponding to them
parameters of DE interactions are also small.
Then, by considering the sum of DE interactions
up to the fourth coordination sphere, we find
$2J_1^{(1)}$$+$$8J_2^{(1)}$$+$$4J_3^{(1)}$$+$$8J_4^{(1)} = 231.50$ meV, which readily reproduces 99 \%
of the kinetic energy $-$$\frac{1}{2}E_{\rm kin} = 234.13$ meV. A very similar conclusion
holds in unrestricted HF and SDMFT calculations.

  Moreover, SDMFT yields very similar parameters of the main DE interactions:
$J_1^{(1)} = 28.71$ meV and $J_2^{(1)} = 19.82$ meV, which are practically undistinguishable from
the ones in DMFT. However, the parameters obtained in the HF calculations
are considerably smaller, especially for the nearest neighbors:
$J_1^{(1)} = 23.75$ meV and $J_2^{(1)} = 19.03$ meV. Such behavior is directly related to the
orbital polarization and additional splitting of the states $2$ and $3$ around the Fermi level (see Fig.~\ref{fig.DOSmodel},
which tend to decrease $|E_{\rm kin}|$ and, therefore, the values of DE interactions.
For instance,
in the case of $J_1^{(1)}$, all transfer integrals (\ref{eqn:t11p}) are diagonal with respect to the orbital
indices. Therefore, as the orbitals $2$ and $3$ become, respectively,
more and less populated in the case of HF calculations (see Table~\ref{tab:n}), the DE interaction $J_1^{(1)}$
will decrease. In the case of $J_2^{(1)}$, the situation is less straightforward, because the transfer
integrals (\ref{eqn:t12}) mix the orbitals $2$ and $3$, counterbalancing the change of the orbital occupations.

  Other contributions to $J_i^{(n)}$ can be found numerically. Particularly, $J_i^{(2)}$ is of the first order of
$[\Delta \hat{\Sigma}(\omega)]^{-1}$. It contains the contributions of superexchange interactions and the
exchange processes between sites separated by two hoppings. In static case, all these parameters
can be expressed via moments of the local density of states \cite{Springer}. However, in dynamic case such
simple relationship does not exist, because of the frequency-dependence of $\hat{\Sigma}$.

\begin{figure}[h!]
\begin{center}
\includegraphics[width=12cm]{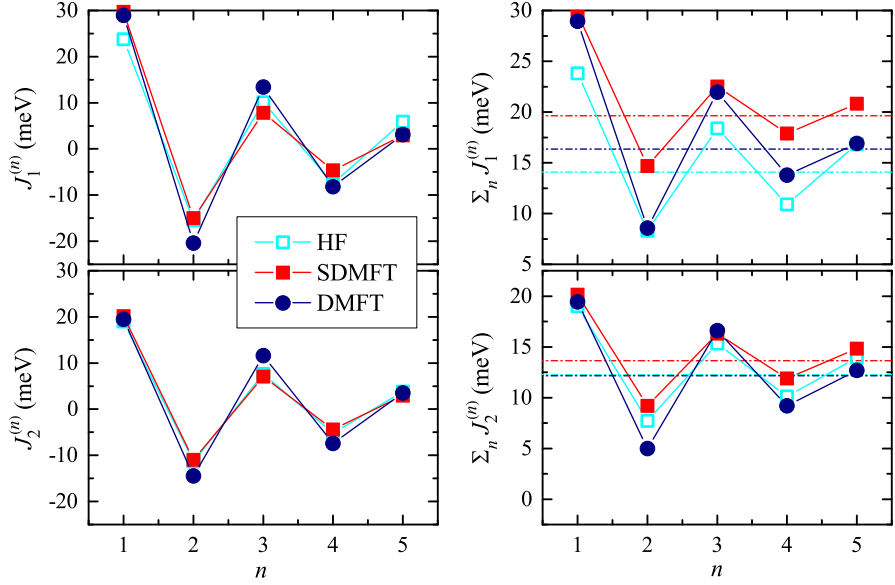}
\end{center}
\caption{\label{fig.Jn}
(Color online) Results of the $(\Delta \hat{\Sigma})^{-1}$ expansion for the nearest-neighbor (top) and
next-nearest-neighbor (bottom) exchange interactions.
The individual contributions, $J_i^{(n)}$, are shown on the left panel and their sum --
on the right panel. The asymptotic values of $J_1$ and $J_2$ are shown by the
dash-dotted lines.
}
\end{figure}

\begin{figure}[h!]
\begin{center}
\includegraphics[width=8cm]{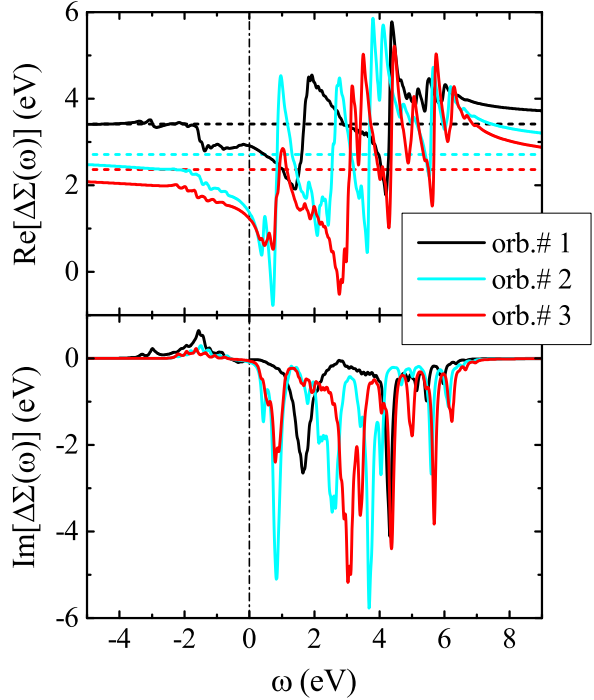}
\end{center}
\caption{\label{fig.DSigma}
(Color online) Frequency dependence of intraatomic spin splitting
$\Delta \hat{\Sigma} (\omega)$ in DMFT. The static limit
$\Delta \hat{\Sigma} (\omega$$\to$$\infty)$ is shown by dashed lines.
The Fermi level is at zero energy (shown by dot-dashed line).
}
\end{figure}

  The results of these calculations are shown in Fig.~\ref{fig.Jn}.
Both $J_1^{(n)}$ and $J_2^{(n)}$ display some characteristic oscillating behavior, where the odd
FM contributions are partially compensated by the even AFM ones. This tendency is observed in all the calculations,
based on the unrestricted HF, SDMFT, and DMFT techniques. The main difference is the convergence of $\sum_n J_i^{(n)}$,
which is noticeably slower in DMFT: the frequency-dependence substantially reduces ${\rm Re}[\Delta \hat{\Sigma}]$
in the occupied part, especially in the region close to the Fermi level (see Fig.~\ref{fig.DSigma}) and,
therefore, slows down the
convergence of the $(\Delta \hat{\Sigma})^{-1}$ expansion. On the other hand, ${\rm Im}[\Delta \hat{\Sigma}]$ is relatively
small in the occupied part and does not play a significant role.
Another important aspect is the cancelation of FM and AFM
contributions to $J_1$ and $J_2$. As was discussed above, the unrestricted HF approach yields somewhat weaker
FM DE contributions
$J_1^{(1)}$ and $J_2^{(1)}$, due to the orbital polarization effects. However, the next AMF contributions
$J_1^{(2)}$ and $J_2^{(2)}$ are also weaker due to the larger spin splitting $\Delta \hat{\Sigma}$
in comparison with DMFT. Thus, the total values of $J_1$ and $J_2$, obtained after summation of all these contributions,
appear to be very close in the case of HF and DMFT.

  The series $\sum_n J_i^{(n)}$ is practically converged for $n=5$, where these sums are close to the
saturated values of $J_1$ and $J_2$. The major FM contribution to $J_1$ and $J_2$ is indeed due to the DE mechanism ($n=1$).
However, this contribution is not the only one and, at least, the $n=3$ term is also very important in
stabilizing the FM ground state. Thus, already from this point of view,
it is not quite right to consider CrO$_2$ as the DE system:
the behavior of $J_1$ and $J_2$ involves
other important mechanisms besides the double exchange and superexchange interactions, which are considered
in the conventional DE model \cite{deGennes,Dagotto}.

\subsection{\label{sec:finite} Magnetic-state dependence of interatomic exchange interactions and Curie temperature}

  Generally speaking, the exchange interactions (\ref{eqn:Jij}) depend on the magnetic state in which they
are calculated. This dependence reflects the change of the electronic structure in different magnetic states and such
information is incorporated in the one-electron Green function
$\hat{G}^{\uparrow, \downarrow}(\omega)$.
The magnetic state-dependence of exchange interactions may have different physical origin.
For instance, it can be the orbital ordering
in insulating \cite{KugelKhomskii} or metallic \cite{PRB01} systems,
or simply the change of the bandwidth in metallic compounds
depending on the magnetic state \cite{deGennes}.
The theory of infinitesimal spin rotations \cite{JHeisenberg,Katsnelson2000}
is more suitable for the
description of effects, which are related to small variations of magnetic moments near the ground state
(for instance, the spin waves). Generally speaking, it is not applicable for the analysis of large perturbations,
such as the spin disorder
near the Curie temperature ($T_{\rm C}$), unless the exchange
interactions do not depend on the magnetic state.

\begin{figure}[h!]
\begin{center}
\includegraphics[width=12cm]{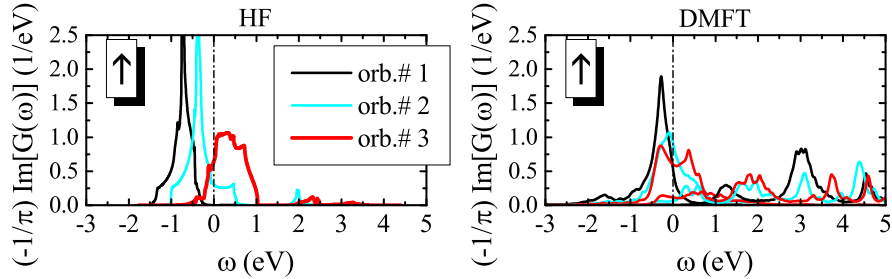}
\end{center}
\caption{\label{fig.DOSmodelA}
(Color online) Partial densities of states as obtained in the unrestricted
Hartree-Fock approach (left) and the dynamical mean-field theory (right)
for the antiferromagnetic state.
The Fermi level is at zero energy (shown by dot-dashed line).
}
\end{figure}

  The comparison of interatomic exchange interactions, calculated in the FM and AFM states
using the theory of infinitesimal spin rotations in the case of
HF and DMFT techniques, is given in Table~\ref{tab:JMState}
(throughout this work we consider the simplest AFM configuration, where the corner and body-centered Cr moments
in the single unit cell are oriented antiferromagnetically).
\begin{table}[h!]
\caption{Parameters of interatomic exchange interactions (in meV) and
corresponding Curie temperature (in Kelvins), obtained using different techniques
and starting conditions, such as
the theory of infinitesimal spin
rotations near the ferromagnetic (F) and antiferromagnetic (A) state in
the frameworks of
unrestricted HF and DMFT methods, as well as the mapping of the total energies
obtained in the self-consistent HF calculations for
the spin-spiral configurations onto Heisenberg model (SCHF).
The notations of parameters $J_i$ are explained in Fig.~\ref{fig.J}.
The dash sign in the row $T_{\rm C}$ means that for the given set of parameters
the ferromagnetic state is unstable.}
\label{tab:JMState}
\begin{ruledtabular}
\begin{tabular}{crrrrr}
 parameter  & \multicolumn{2}{c}{HF} & \multicolumn{2}{c}{DMFT} & SCHF \\
 \cline{2-3} \cline{4-5}
                & F~~ & A~~ & F~~ & A~~ \\
\hline
  $J_1$     & $14.06$ & $23.77$ & $16.35$ & $18.14$ & $11.00$ \\
  $J_2$     & $12.26$ & $14.91$ & $12.14$ &  $6.73$ & $15.43$ \\
  $J_3$     &  $1.16$ &  $0.25$ &  $0.60$ &  $0.21$ &  $2.90$ \\
  $J_4$     &  $0.96$ &  $1.39$ &  $0.35$ & $-1.08$ &  $1.43$ \\
  $J_5$     & $-0.39$ &  $1.39$ & $-1.15$ & $-2.66$ &  $0.10$ \\
  $J_6$     & $-1.87$ & $-0.14$ & $-1.85$ &  $0.22$ & $-1.36$ \\
  $J_7^<$   & $-1.21$ & $-6.21$ & $-2.58$ & $-2.68$ & $-4.13$ \\
  $J_7^>$   & $-3.26$ & $-7.99$ & $-4.19$ & $-5.03$ & $-4.13$ \\
  $J_8^<$   & $-0.31$ & $-0.94$ & $-0.94$ & $-3.57$ & $-1.41$ \\
  $J_8^>$   & $-0.46$ & $-3.05$ & $-2.44$ & $-1.60$ & $-1.41$ \\
\hline
$T_{\rm C}$ & $581$   &  $-$~~  & $-$~~   & $-$~~   & $684$   \\
\end{tabular}
\end{ruledtabular}
\end{table}
One can clearly see that the exchange interactions are quite sensitive to the magnetic state
in which they are calculated. Generally, the AFM structure remains unstable and is not the
ground state of CrO$_2$. Nevertheless, the AFM spin alignment tends to reconstruct the electronic structure
(Fig.~\ref{fig.DOSmodelA}) in such a way as
to additionally stabilize the FM interactions $J_1$ in the NN
ferromagnetically coupled bond. Moreover, in DMFT,
the FM interactions $J_2$ in the
antiferromagnetically coupled next-NN bond is strongly reduced, that also works in the direction of
stabilizing the AFM state (and destabilizing the FM one).
Similar tendency holds for longer-range interactions. Thus, if one tries to use the parameters obtained
in the AFM state in order to describe the FM state, one can easily find that this FM state will be unstable,
even in the unrestricted HF approach. Perhaps, this was an extreme example, and below we will
consider a more realistic strategy for the evaluation of $T_{\rm C}$.

  Taking into consideration the strong magnetic state-dependence of exchange interactions,
we tried to go beyond the theory of infinitesimal spin rotations and evaluated the exchange
interactions using results of self-consistent total energy calculations for spin-spiral configurations
with arbitrary wavevectors ${\bf q}$. Namely, using generalized Bloch theorem \cite{Sandratskii}, we performed the
unrestricted HF calculations for spin-spiral configurations, where the directions
of magnetic moments varied as
$$
\boldsymbol{e}_{\boldsymbol{\tau}+{\bf R}} = \left(
\begin{array}{c}
\cos (\boldsymbol{\tau}+{\bf R}) \cdot {\bf q} \\
\sin (\boldsymbol{\tau}+{\bf R}) \cdot {\bf q} \\
0 \\
\end{array}
\right),
$$
calculated the total energy ($E_{\bf q}$) for each ${\bf q}$, and evaluated the exchange interactions
as the Fourier transform of $E_{\bf q}$.
The results are also listed in Table~\ref{tab:JMState}, in the column `SCHF'.
Particularly, we expected that the exchange interactions, obtained by mapping the total energies
of the spin-spiral configurations onto the Heisenberg model, should provide a good estimate
for $T_{\rm C}$.
The latter was evaluated using Tyablikov's random-phase approximation \cite{spinRPA}.
The results are also listed in Table~\ref{tab:JMState} and can be summarized as follows:
as long as we use the unrestricted HF approximation, supplemented either with the
theory of infinitesimal spin rotations near the FM state or with the self-consistent
spin-spiral calculations for finite ${\bf q}$'s, running through the first Brillouin zone,
$T_{\rm C}$ is even overestimated in comparison with the experimental data, meaning that
the FM state is indeed very robust. However, when we switch to more rigorous DMFT technique,
the FM state appears to be unstable because of the longer-range AFM interactions
(and any numerical estimates of $T_{\rm C}$ in this case become meaningless).
This is a very serious problem, which we will discuss in details in the next section.

\subsection{\label{sec:JLR} Long-range interactions and stability of the ferromagnetic state}

  In Sec.~\ref{sec:DE}, we have seen that, as far as the NN and next-NN interactions are concerned,
unrestricted HF and DMFT techniques very similar results. Nevertheless, there is an important
difference in the behavior of longer-range interactions, which has fundamental consequences.
Since the frequency-dependence reduces intraatomic spin splitting ${\rm Re}[\Delta \hat{\Sigma} (\omega)]$ near the
Fermi level, the series (\ref{eqn:nseries}) converges somewhat slower in the case of DMFT.
Besides oscillating behavior depicted in Fig.~\ref{fig.Jn}, smaller ${\rm Re}[\Delta \hat{\Sigma} (\omega)]$ is responsible
for larger spacial extension of the exchange interactions. This can be directly seen from the construction
$\left( [\Delta \hat{\Sigma}(\omega) \hat{G}^{\uparrow}(\omega)]^{-1} \right)^n_{i0}$ in Eq.~(\ref{eqn:nterm}):
since $[\hat{G}^{\uparrow}(\omega)]^{-1}_{ij} = \hat{t}_{ij}$ for $i$$\ne$$j$, and the transfer integrals
are typically restricted by only few coordination spheres, the $n$-order term will include the processes,
which connect two remote sites $0$ and $i$ by $n$ sequential hoppings between nearest or next nearest
neighbors. Obviously, such contributions will be stronger for smaller ${\rm Re}[\Delta \hat{\Sigma} (\omega)]$.
Moreover, the number of nodes of the integrand in Eq.~(\ref{eqn:nterm}) increases with the
distance between $0$ and $i$ \cite{Heine1,Heine2}. Therefore, it is possible that some of these
long-range interactions can easily become antiferromagnetic. Such behavior is clearly seen in
Table~\ref{tab:J} and Fig.~\ref{fig.J}: besides FM interactions, there are several relatively strong
AFM interactions, connecting the sites in the 5-8 coordination spheres. These interactions are
stronger in DMFT because of smaller spin splitting ${\rm Re}[\Delta \hat{\Sigma} (\omega)]$.

\begin{figure}[h!]
\begin{center}
\includegraphics[width=12cm]{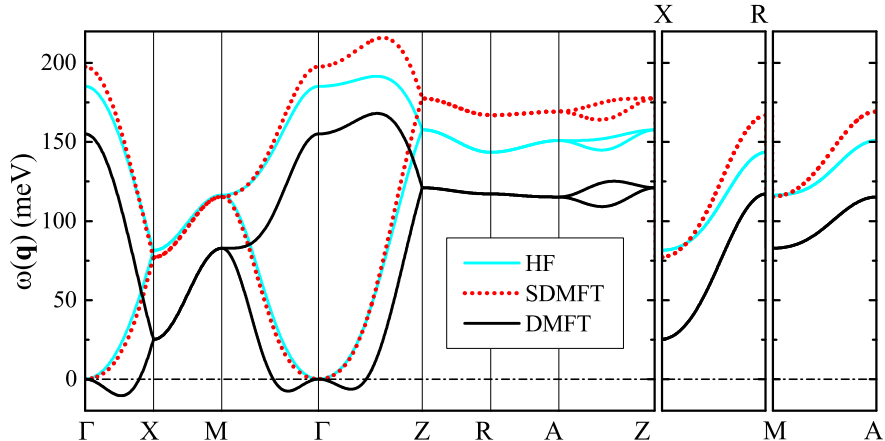}
\end{center}
\caption{\label{fig.SW}
(Color online)
Results of calculations of the spin-wave dispersion with the parameters, obtained in
the theory of infinitesimal spin rotations in the case of HF,
SDMFT, and DMFT techniques.
Notations of the high-symmetry points of the Brillouin zone are taken from \cite{BradleyCracknell}.
}
\end{figure}

  The appearance of AFM interactions naturally rises the question
about stability of the FM state and whether it
is indeed the magnetic ground state of the considered model.
In order to investigate this problem, we evaluate the spin-wave dispersion,
$\omega({\bf q})$,
using the interatomic exchange interactions obtained in the theory of infinitesimal spin rotations.
In the $P4_2/mnm$ structure, containing two magnetic sublattices, $\omega({\bf q})$ can be obtained
from the diagonalization of the $2$$\times$$2$ matrix (for $S$$=$$1$):
$$
\hat{\Omega}({\bf q}) =
\left(
\begin{array}{cc}
J_{11}({\bf q}) - J_0 & J_{12}({\bf q}) \\
J_{21}({\bf q}) & J_{22}({\bf q}) - J_0 \\
\end{array}
\right),
$$
where $J_{\alpha \beta}({\bf q})$ is the Fourier image of magnetic interactions between
sublattices $\alpha$ and $\beta$, and $J_0 = J_{11}(0)+J_{12}(0)$. In principle,
due to the symmetry properties, $J_{22}({\bf q})$ can be related to
$J_{11}({\bf q}^*)$ in some other ${\bf q}$-point.
The same holds for $J_{12}({\bf q})$ and $J_{21}({\bf q}^*)$. The results of these calculations
are shown in Fig.~\ref{fig.SW}.
The negative spin-wave frequencies signal that the FM state is unstable.
One can clear see that as long as we use the static HF and SDMFT techniques,
there is no problem with the stability of the FM state, and
$T_{\rm C}$
is even overestimated in comparison with the experimental data (Table~\ref{tab:JMState}). Thus, one could naively think that
the FM state is very robust. Nevertheless, in DMFT, which is definitely the most rigorous approach among
the considered ones, the FM state appears to be unstable. This instability occurs along
three high symmetry directions of the Brillouin zone
($\Gamma$-${\rm X}$, $\Gamma$-${\rm M}$, and $\Gamma$-${\rm Z}$). This is a very serious problem,
meaning that there should be additional factors, which are not taken into account in the low-energy electron model
for the $t_{2g}$ bands and
which stabilize the FM state. This problem will be studied in the next section.

\subsection{\label{sec:Oband} Direct exchange interactions and contributions of the oxygen states}

  In this section, we evaluate the change of magnetic energy, caused by the polarization of the O $2p$ band
and other contributions, which are not taken into account in the minimal model for the $t_{2g}$ bands.

  For these purposes, after solution of the low-energy model in DMFT,
we go back from the Wannier basis $\{ \phi_{\tau a} \}$
of the model to the original LMTO basis $\{ \chi_{\upsilon b} \}$:
\begin{equation}
\phi_{\tau a}(\textbf{r} - \boldsymbol{\tau}) = \sum_{\upsilon b}
q_{\tau a}^{\upsilon b} \chi_{\upsilon b} (\textbf{r} - \boldsymbol{\upsilon}),
\label{eqn:overLMTO}
\end{equation}
and construct the spin magnetization density,
$m({\bf r}) = n_{\uparrow}({\bf r})$$-$$n_{\downarrow}({\bf r})$,
associated with the Cr $t_{2g}$ band.
This $m({\bf r})$ has major contributions at the Cr sites as well as some hybridization-induced contribution
at the oxygen sites. Following the philosophy of the low-energy model \cite{review2008},
the interaction of $m({\bf r})$
with the rest of the electronic states should be well described
already at the LSDA level. Therefore, our strategy
is to evaluate, in LSDA, the exchange-correlation (xc) field $b({\bf r}) = v_{\downarrow}({\bf r})$$-$$v_{\uparrow}({\bf r})$
($v_{\uparrow,\downarrow}$ being the xc potential in LSDA),
which is induced by $m({\bf r})$ and polarizes the O $2p$ band,
and find the self-consistent change of $m({\bf r})$ and $b({\bf r})$, caused by
the interaction between $t_{2g}$ and O $2p$ bands.
For these purposes, it is convenient to use the
self-consistent linear response (SCLR) theory \cite{SCLR}.
For simplicity, let us consider the discrete lattice model and assume that all weights of $m({\bf r})$
are concentrated in the lattice points:
$m({\bf r}) = \sum_{\upsilon} m_{\upsilon} \delta (\textbf{r} - \boldsymbol{\upsilon})$,
where $m_{\upsilon}$ is the local magnetic moment at the site $\upsilon$. Furthermore,
we recall that LSDA is conceptually close to the Stoner model, where
the xc energy is given by \cite{Gunnarsson}:
\begin{equation}
E_{\rm xc} = -\frac{1}{4} \sum_{\upsilon} I_{\upsilon} m_{\upsilon}^2.
\label{eqn:Stoner}
\end{equation}
In practical calculations, the parameters $\{ I_{\upsilon} \}$ can be found using the values of
intraatomic spin splitting and local magnetic moments in LSDA. Meanwhile the intraatomic spin splitting itself
can be obtained using LMTO parameters of the centers of gravity for
the $\uparrow$- and $\downarrow$-spin states \cite{LMTO2},
which yields
$I_{\rm Cr} = 0.98$ eV and $I_{\rm O} = 1.68$ eV.

  Then, the self-consistent field can be found as
$$
\vec{b} = \left[ 1 + \hat{\cal I} \hat{\cal R} \right]^{-1} \vec{b}^{\,0},
$$
where we have introduced the vector $\vec{b} \equiv [ b_{\upsilon} ]$ and
the tensors
$\hat{\cal I} = [ I_{\upsilon} \delta_{\upsilon \upsilon'}]$
and $\hat{\cal R} = [ {\cal R}_{\upsilon \upsilon'} ] $. In this equation,
$\vec{b}^{\,0} = \hat{\cal I} \vec{m}$ is the xc field induced by the Cr $t_{2g}$ band, and the response tensor $\hat{\cal R}$
is obtained in the first order perturbation theory for the wavefunctions, starting from the nonmagnetic LDA band structure:
\begin{equation}
{\cal R}_{\upsilon \upsilon'} = \sum_{ab} \sum_n^{\rm occ} \sum_{n'}^{\rm unocc} \sum_{\bf k}^{\rm BZ}
\left\{
\frac{(C_{n {\bf k}}^{\upsilon a})^* C_{n' {\bf k}}^{\upsilon a}
(C_{n' {\bf k}}^{\upsilon' b})^* C_{n {\bf k}}^{\upsilon' b}}
{\varepsilon_{n {\bf k}} - \varepsilon_{n' {\bf k}}}
+
{\rm c.\,c.}
\right\},
\label{eqn:Rtensor}
\end{equation}
where $\{ C_{n {\bf k}}^{\upsilon a} \}$ are the coefficients of the expansion of the LDA wavefunctions
over LMTO's, $\{ \varepsilon_{n {\bf k}} \}$ are the LDA eigenvalues, and ${\bf k}$ runs over the
first Brillouin zone (BZ). Moreover, similar to the constrained RPA \cite{Ferdi04},
we have to exclude from Eq.~(\ref{eqn:Rtensor}) contributions, where both indexes $n$ and $n'$ belong to the Cr $t_{2g}$ band.
In the perturbation theory, such
terms describe the change of the magnetization in the $t_{2g}$ band, which are caused by the LSDA potential.
However, in the low-energy model, this part is replaced by the more rigorous DMFT solution
with the screened Coulomb interactions. Therefore,
in order avoid the double counting, such contributions should be excluded
in the process of SCLR calculations. In practice, $n$ runs over the occupied O $2p$ bands and
$n'$ runs over the unoccupied Cr $t_{2g}$ and $e_g$ bands.

  Once the self-consistent field $\vec{b}$ is known, the change of $\vec{m}$ and $\vec{b}$,
caused by the polarization of the oxygen band, can be found as
$\delta \vec{m} = - \hat{\cal R} \vec{b}$
and $\delta \vec{b} = \hat{\cal I} \delta \vec{m}$, respectively.
Since the O $2p$ band is occupied, the net change of magnetic moment will vanish:
$\sum_{\upsilon} \delta m_{\upsilon} = 0$, irrespectively on the type of the magnetic order.
Nevertheless, the individual moments $\delta m_{\upsilon}$ can be finite and contribute to the total energy.
The corresponding energy change, caused by the magnetic
polarization of the oxygen band, consists of two parts:
$\delta E^{\rm pol} = \delta E_{\rm Cr-O}^{\rm pol} + \delta E_{\rm O}^{\rm pol}$,
where $\delta E_{\rm Cr-O}^{\rm pol} = -\frac{1}{2} \delta \vec{m}^T \hat{\cal I} \vec{m}$
is the interaction of $\delta m_{\upsilon}$ with the ``external'' xc field, created by the Cr $t_{2g}$ band,
and $\delta E_{\rm O}^{\rm pol}$ is the energy change caused by $\delta \vec{m}$ in the O $2p$ band.
It also consists of two parts:
$\delta E_{\rm O}^{\rm pol} = \delta E_{\rm sp} + \delta E_{\rm dc}$,
where $\delta E_{\rm sp}$ is the single-particle energy,
which can be found in the second order of $\delta \vec{b}$ as
$\delta E_{\rm sp} = \frac{1}{4} \delta \vec{b}^T \hat{\cal R} \vec{b}$ \cite{SCLR}, and
$\delta E_{\rm dc} = \frac{1}{4} \delta \vec{m}^T \hat{\cal I} \delta \vec{m}$ is the double-counting energy,
where $\delta \vec{m}^T$ is the row vector, corresponding to the column vector $\delta \vec{m}$.
In all these calculations,
it is assumed that the magnetic energy of the $t_{2g}$ band itself is
described by DMFT.

  The polarization energy $\delta E^{\rm pol}$ may have different values in the case of the FM and AFM alignment of spins
and, thus,
contributes to
interatomic exchange interactions.
Below we evaluate this effect in CrO$_2$.
The magnetic moments are listed in Table~\ref{tab:A1} and the energies are in
Table~\ref{tab:A2}.
\begin{table}[h!]
\caption{The values of local magnetic moments at the chromium and oxygen sites $\{ m_{\upsilon} \}$ as
obtained in DMFT calculations for the
isolated $t_{2g}$ band in the case of
ferromagnetic (F) and antiferromagnetic (A) alignment of Cr spins, and
the moments $\{ \delta m_{\upsilon} \}$, caused by the polarization of the O $2p$ band.
All values are in $\mu_{\rm B}$.}
\label{tab:A1}
\begin{ruledtabular}
\begin{tabular}{lcccc}
 & \multicolumn{2}{c}{F} & \multicolumn{2}{c}{A} \\
 \cline{2-3} \cline{4-5}
                & $m_{\upsilon}$  & $\delta m_{\upsilon}$
                & $m_{\upsilon}$  & $\delta m_{\upsilon}$ \\
\hline
Cr & $1.628$ & $\phantom{-}0.594$  & $1.392$ & $\phantom{-}0.584$  \\
O  & $0.134$ & $-0.297$            & $0.029$     & $-0.089$      \\
\end{tabular}
\end{ruledtabular}
\end{table}
\begin{table}[h!]
\caption{The energy changes (in meV per one formula unit), caused by the magnetic polarization
of the O $2p$ band in the ferromagnetic (F) and antiferromagnetic (A) states:
the interaction energy between Cr $t_{2g}$ and O $2p$ bands ($\delta E_{\rm Cr-O}^{\rm pol}$),
the magnetic energy in the O $2p$ band ($\delta E_{\rm O}^{\rm pol}$),
and the total energy ($\delta E^{\rm pol} = \delta E_{\rm Co-O}^{\rm pol} + \delta E_{\rm O}^{\rm pol}$).
All values were derived using DMFT magnetization density for the $t_{2g}$ band.
}
\label{tab:A2}
\begin{ruledtabular}
\begin{tabular}{lrr}
    & F~~~   &   A~~~ \\
\hline
  $\delta E_{\rm Cr-O}^{\rm pol}$                  &  $-449.27$ &  $-434.67$ \\
  $\delta E_{\rm O}^{\rm pol}$                     &  $99.04$   &  $66.42$ \\
  $\delta E^{\rm pol}$                             &  $-350.24$ &  $-368.24$ \\
\end{tabular}
\end{ruledtabular}
\end{table}

 The spin moments $m_{\upsilon}$ are redistributed between Cr and oxygen sites.
As expected for the FM state,
the total moment is $m_{\rm Cr} + 2m_{\rm O} = 1.9$ $\mu_{\rm B}$,
which is totally consistent with the value obtained in the Wannier basis (see Table~\ref{tab:n}).
The small deviation from $2$ $\mu_{\rm B}$ is caused by nonquasiparticle $\downarrow$-spin states near the Fermi level.
The moments $m_{\upsilon}$ and $\delta m_{\upsilon}$ are parallel at the Cr sites
and antiparallel at the oxygen sites. This tendency is consistent with results of first-principles calculations
and can be deduced from the analysis of hybridization between Cr $3d$ and O $2p$ states \cite{JPSJ}. Therefore,
the negative sign of $\delta E_{\rm Cr-O}$ is due to the contributions of the Cr sites,
which are partly compensated by positive contributions of the oxygen sites. The absolute value of $\delta E_{\rm Cr-O}$
is larger in the FM state, mainly because $m_{\rm Cr}$ and $\delta m_{\rm Cr}$ are larger.
Thus, the Cr-O interaction additionally stabilizes the FM state. The contribution of the O $2p$ band to the magnetic energy is positive.
This is because the O $2p$ band itself does not favor the magnetism and any magnetic polarization of this band
will increase the total energy.
This also explains why
$\delta E_{\rm O}$ is smaller in the AFM state: the magnetic moments $\delta m_{\upsilon}$ are smaller
and, therefore, the magnetic perturbation of the O $2p$ band is also smaller. In CrO$_2$, the
second effect ($\delta E_{\rm O}$)
dominates and the polarization of the O $2p$ slightly favors the AFM alignment.
The corresponding energy difference between FM and AFM states,
$\Delta E^{\rm pol} = \delta E^{\rm pol}({\rm F})$$-$$\delta E^{\rm pol}({\rm A})$,
is about $18$ meV per one formula unit.

  Another contribution to the magnetic energy is related to
the direct interactions between Wannier functions in the $t_{2g}$ bands \cite{Ku,Mazurenko},
which are centered at different Cr sites.
They are not taken into account in the low-energy model,
because the latter treats only on-site Coulomb and exchange interactions.
Nevertheless, these interactions can be evaluated in LSDA. First, let us evaluate the difference of LSDA xc energies
between FM and AFM states in the $t_{2g}$ band,
$\Delta E_{\rm xc} = \delta E_{\rm xc}({\rm F})$$-$$\delta E_{\rm xc}({\rm A})$,
using the values of magnetic moments $\{ m_{\upsilon} \}$ from
Table~\ref{tab:A1} and Eq.~(\ref{eqn:Stoner}) for $E_{\rm xc}$.
This yields $\Delta E_{\rm xc} =  -$$199.65$ meV per one formula unit, where the main contribution
(about 93\%)
comes from the Cr sites.
This energy difference favors the FM alignment.
Then, we note that the xc interaction between Wannier orbitals centered at the same Cr site
is already taken into account in the low-energy model in the framework of
DMFT. Therefore, we should subtract this on-site ``self-interaction'' (SI) part from the LSDA xc energy difference.
This can be done as follow. Using the spin magnetization matrix
$$
\hat{\cal M}_{\tau} \equiv [{\cal M}_{\tau}^{ab}]
= -\frac{1}{\pi} {\rm Im} \int_{- \infty}^{\varepsilon_{\rm F}} d \omega \, \left[
\hat{G}_{\tau \tau}^{\uparrow}(\omega) - \hat{G}_{\tau \tau}^{\downarrow}(\omega)
\right],
$$
obtained in the Wannier basis at the Cr site $\tau$,
and the expansion (\ref{eqn:overLMTO}) over LMTO's, we evaluate magnetic moments, that are produced by
$\hat{\cal M}_{\tau}$ at the central and neighboring to it sites $\upsilon$:
$$
\bar{m}_{\upsilon} = \sum_{abc} \left( q_{\tau a}^{\upsilon c} \right)^* {\cal M}_{\tau}^{ab} q_{\tau b}^{\upsilon c}.
$$
The difference between $\bar{m}_{\upsilon}$ and $m_{\upsilon}$ is that $\bar{m}_{\upsilon}$ is the
contribution of the \textit{single Cr site} $\tau$ to the magnetic moment at the site $\upsilon$,
while $m_{\upsilon}$ takes into account the contributions of \textit{all sites of the Cr lattice}.
Therefore, $\bar{m}_{\upsilon}$ at the central site $\tau$ is substantially smaller than $m_{\upsilon}$
($\bar{m}_{\upsilon}$$=$ $1.169$ and $1.099$ $\mu_{\rm B}$ in the FM and AFM state, respectively).
The total moment $\sum_{\upsilon} \bar{m}_{\upsilon}$ in the FM state is only $1.543$ $\mu_{\rm B}$,
which also substantially deviates from $\sum_{\upsilon} m_{\upsilon} = 1.9$ $\mu_{\rm B}$. Then, we evaluate the
SI energy, which is also given by (\ref{eqn:Stoner}), but after replacing $\{ m_{\upsilon} \}$
by $\{ \bar{m}_{\upsilon} \}$. This yields the following energy difference between FM and AFM states:
$\Delta E_{\rm SI} \equiv \delta E_{\rm SI}({\rm F})$$-$$\delta E_{\rm SI}({\rm A}) = -$$39.19$ meV per one formula unit.
Therefore, by subtracting the SI term, we will additionally shift the energy balance in the favor
of antiferromagnetism.

\begin{figure}[h!]
\begin{center}
\includegraphics[width=8cm]{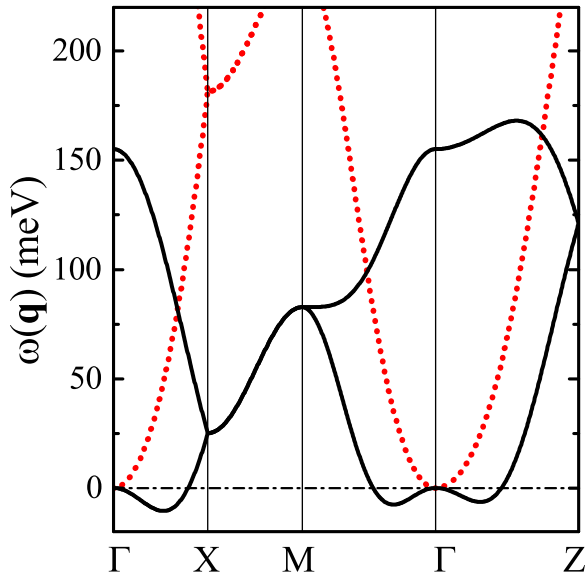}
\end{center}
\caption{\label{fig.SWm}
(Color online)
Results of calculations of the spin-wave dispersion with DMFT parameters obtained
for the isolated $t_{2g}$ band (solid line) and after taking into account the
additional ferromagnetic contribution $\Delta J_2 = 17.81$ meV, arising from magnetic polarization
of the oxygen band and direct exchange interactions in the $t_{2g}$ band (dotted line).
Notations of the high-symmetry points of the Brillouin zone are taken from \cite{BradleyCracknell}.
}
\end{figure}

  Thus, by combining all the contributions, the total energy difference
$\Delta E = \Delta E^{\rm pol} + \Delta E_{\rm xc} - \Delta E_{\rm SI}$ is about $-$$142.46$ per one formula unit.
By mapping this total energy difference onto the Heisenberg model and assuming that it contributes only to the
next-NN interactions, one can find the following correction to this interaction, arising from the polarization of the
oxygen band and direct exchange interactions in the $t_{2g}$ band: $\Delta J_2 \equiv -$$\Delta E/8 = 17.81$ meV.
The spin-wave dispersion, which takes into account the additional FM contribution $\Delta J_2$ is plotted
in Fig.~\ref{fig.SWm} in comparison with results of regular DMFT calculations for the isolated $t_{2g}$ band.
One can clear see that all $\omega({\bf q})$ in this case become nonnegative and the FM state is stable. Thus,
the magnetic polarization of the oxygen band and direct exchange interactions in the $t_{2g}$ band
play a very important role in the stability of the FM state in CrO$_2$.

\subsection{\label{sec:dcorr} Optimized effective potential method and importance of dynamic correlations}
In this section, we discuss results of the optimized effective potential method (OEP) \cite{TalmanShadwick,KotaniAkai,Kotani,EngelSchmid,GraboGross},
which we consider mainly for pedagogical purposes, in order to emphasize the importance of careful treatment of the correlation effects.
OEP is a numerical realization of the Kohn-Sham density functional theory \cite{KohnSham},
where
\begin{enumerate}
\item
the one-electron band structure is obtained from solution of
Schr\"odinger equations with some effective static local potential $\hat{v}$:
\begin{equation}
\left( \hat{t}_{\bf k} + \hat{v} \right) | c_{n {\bf k}} \rangle =
\varepsilon_{n {\bf k}} | c_{n {\bf k}} \rangle;
\label{eqn:KS}
\end{equation}
\item
the obtained band structure is used to calculate the total energy
\begin{equation}
E = E_{\rm kin} + E_{\rm C} + E_{\rm X} + E_{\rm corr},
\label{eqn:Etot}
\end{equation}
consisting of kinetic ($E_{\rm kin}$, which also includes the energy of crystal-field splitting),
Coulomb ($E_{\rm C}$), exchange ($E_{\rm X}$),
and correlation ($E_{\rm corr}$) parts;
\item
the parameters of effective potential $\hat{v}$ are found numerically, so to minimize the
total energy (\ref{eqn:Etot}).
\end{enumerate}
Thus, the OEP method provides some alternative possibility for the construction of static potential,
which, in addition to the standard Coulomb and exchange contribution, includes the effect of
correlation interactions and, in this sense, can be regarded as a step beyond the HF approximation.

  This aforementioned OEP procedure was implemented for the solution of the low-energy model
for the $t_{2g}$ band \cite{OEP2011}.
Here, we assume that the one-electron band structure is half-metallic and all minority-spin states are unoccupied.
Therefore, we drop the spin indices, but keep in mind that both the potential and electronic structure
are referred to the $\uparrow$-spin states. Then, because of the symmetry, the potential matrix is diagonal
$\hat{v} = \left[ v_{ab} \delta_{\tau \tau'} \delta_{ab} \right]$ and does not depend on the indices
$\tau =$ $1$ or $2$ of the Cr-atoms in the primitive cell.
Therefore, the effective potential has only two
independent parameters (apart from the constant energy shift):
$\Delta_{2-1} = v_{22}$$-$$v_{11}$ and $\Delta_{3-2} = v_{33}$$-$$v_{22}$.
Note also that the eigenvector $| c_{n {\bf k}} \rangle$ in
Eq.~(\ref{eqn:KS}) is the row-vector of the form $| c_{n {\bf k}} \rangle = [c_{n {\bf k}}^{a \tau}]$.

  The correlation energy can be evaluated in
RPA as \cite{Pines,FerdiPRL2002}:
\begin{equation}
E_{\rm corr} = \frac{1}{4 \pi} \sum_{\bf q} \int_0^{\infty} d \omega \,
{\rm Tr} \left\{
\ln \left[1-\hat{P}(i\omega,{\bf q})\hat{U}\right]
\left[1-\hat{U}\hat{P}(i\omega,{\bf q})\right]
+ 2 \hat{P}(i\omega,{\bf q})\hat{U}
\right\},
\label{eqn:EcRPA}
\end{equation}
where $\hat{U}$ is the matrix $\left[ U_{abcd} \right]$ of the on-site Coulomb interaction and
$\hat{P} = \left[ P^{\tau \tau'}_{abcd} \right]$
is the polarization in the imaginary frequency:
\begin{equation}
P^{\tau \tau'}_{abcd}(i\omega,{\bf q}) = \sum_n^{\rm occ} \sum_{n'}^{\rm unocc}
\sum_{\bf k}
\frac{2 ( \varepsilon_{n {\bf k}} - \varepsilon_{n' {\bf k}+{\bf q}} )}
{\omega^2 + ( \varepsilon_{n {\bf k}} - \varepsilon_{n' {\bf k}+{\bf q}} )^2}
c_{n' {\bf k}+{\bf q}}^{a \tau *} c_{n {\bf k}}^{b \tau}
c_{n {\bf k}}^{c \tau' *} c_{n' {\bf k}+{\bf q}}^{d \tau'}.
\label{eqn:Polarization}
\end{equation}
The matrix multiplication in Eq.~(\ref{eqn:EcRPA}) implies the summation over
two intermediate orbital indices: $(\hat{U}\hat{P})^{\tau \tau'}_{abcd} \equiv
\sum_{ef} U_{abef} P^{\tau \tau'}_{efcd}$ and the $\omega$-integration has been
performed using 10-points Gaussian quadrature method \cite{FerdiPRL2002}.

  By applying this OEP approach, we expected that the correlation effects, beyond the HF approximation,
will reduce the orbital polarization and yield an improved description, at least for the
majority-spin states and DE interactions. Moreover, we expected RPA to
work reasonably well for metallic systems, such as CrO$_2$.
Since the RPA total energy of HM systems
does not depend on the position of unoccupied $\downarrow$-spin states,
we cannot easily determine in the framework of this method the spin-splitting
$\Delta \hat{\Sigma}$ and the parameters of exchange
interactions, which depend on $\Delta \hat{\Sigma}$ and $\hat{G}^{\downarrow}(\omega)$.
Nevertheless, at least we should be able to evaluate
the DE interactions, which do not depend on $\Delta \hat{\Sigma}$.

\begin{figure}[h!]
\begin{center}
\includegraphics[width=12cm]{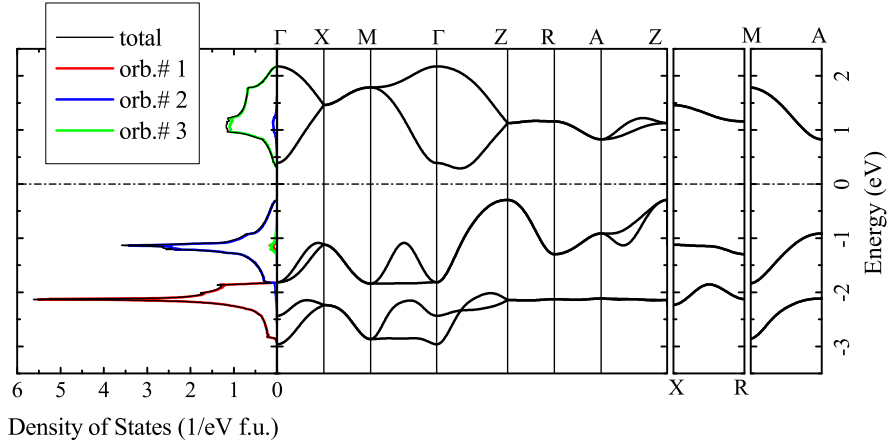}
\end{center}
\caption{\label{fig.OEP}
(Color online) Electronic band structure of CrO$_2$, obtained in the OEP approach: (Left) Total and
partial densities of states of three $t_{2g}$ orbitals and (Right) band dispersion along high-symmetry
directions of the Brillouin zone
(notations of the high-symmetry points are taken from \cite{BradleyCracknell}).
}
\end{figure}

\begin{figure}[h!]
\begin{center}
\includegraphics[width=6cm]{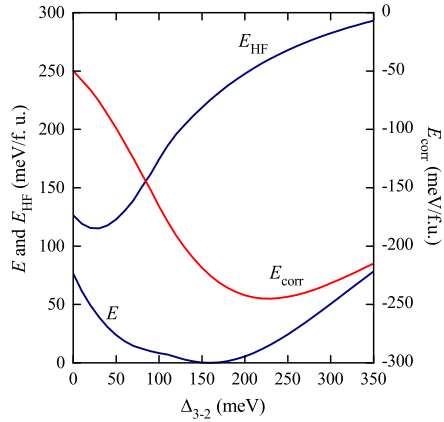}
\end{center}
\caption{\label{fig.EOEP}
(Color online) Results of energy minimization in the OEP method versus the splitting in the potential
between atomic levels 3 and 2: the Hartree-Fock part of the energy
$E_{\rm HF} = E_{\rm kin} + E_{\rm C} + E_{\rm X}$, the correlation energy $E_{\rm corr}$,
and the total energy $E = E_{\rm HF} + E_{\rm corr}$.
}
\end{figure}

  However, somewhat surprisingly, we have obtained very curious, but unphysical result:
the correlation interactions, treated in RPA with the \textit{static} effective potential, tend to additionally
stabilize the orbital ordering and \textit{increase} the orbital polarization,
leading to the \textit{insulating} solution, which is shown in Fig.~\ref{fig.OEP}.
The reason for such unphysical behavior is that, even when the HF energy $E_{\rm HF} = E_{\rm kin}$$+$$E_{\rm C}$$+$$E_{\rm X}$
reaches its minimum, $E_{\rm corr}$ continues to decrease as a function of $\Delta_{3-2}$ (Fig.~\ref{fig.EOEP}).
Obviously, $E_{\rm corr}$ decreases when the polarization decreases (note that $\hat{P}$
is the negative-defined matrix in the imaginary frequency). Then, there are two competing effects.
On the one hand, the additional
splitting of orbitals $2$ and $3$ across the Fermi energy is expected to suppress the correlation interactions.
This is rather general property of correlation energy,
which follows from the perturbation theory analysis \cite{Callaway}.
If there were no transfer integrals, connecting the orbitals $2$ and $3$, the effect of $\Delta_{3-2}$
would be equivalent to the scissors operator and the behavior of $E_{\rm corr}$ would be totally described
by the above mentioned mechanism (which indeed dominates for large $\Delta_{3-2}$).
Nevertheless, the strong hybridization between orbitals $2$ and $3$ [see Eq.~(\ref{eqn:t12})] may change
this canonical behavior. First, we note that, in order to produce a large contribution to $E_{\rm corr}$,
one should activate the channels involving the large Coulomb matrix elements $U_{aacc}$
(where $a$ and $c$ are $2$ or $3$). This can be done only if the polarization matrix has sizable elements
of the same $P_{aacc}$ type [see Eq.~(\ref{eqn:EcRPA})].
Such matrix elements are indeed produced by the hybridization effects in the insulating state [see Eq.~(\ref{eqn:Polarization})]:
because of the hybridization, the orbital $3$ may have a substantial weight in the occupied part of the spectrum
(so as the orbital $2$ in the unoccupied part), yielding finite matrix elements $P_{aacc}$.
Thus, we believe that the decreasing of $E_{\rm corr}$ in Fig.~\ref{fig.EOEP} is a specific property of CrO$_2$
and related to the strong hybridization between occupied and unoccupied orbitals in the
orbitally polarized state. Nevertheless, such a behavior is, of course, unphysical and this example
demonstrates again the importance of explicit consideration of dynamic correlations.

  In order to conclude this section, let us
evaluate the consequences of the exaggerated orbital polarization and gap opening
on the DE interactions. The kinetic energy (without the energy of the crystal field splitting),
obtained in the OEP method, is only $-168.57$ meV per formula unit
and corresponding parameters of DE interactions can be estimated as $J^{(1)}_1 = 3.80$ meV and $J^{(1)}_2 = 8.74$ meV.
Thus, as expected, the DE interactions are strongly underestimated in the OEP approach.

\section{\label{sec:Summary} Summary and conclusions}

  We have presented detailed theoretical analysis of interatomic exchange interaction in CrO$_2$,
which was based on realistic low-energy model, derived from the first-principles electronic structure
calculations, and have involved various techniques for treating electron correlations in the narrow $t_{2g}$ band,
ranging from the static Hartree-Fock approximation to the dynamical mean-field theory. Such analysis allowed us
to elucidate different contributions to the exchange couplings and understand the origin of these
contributions on the microscopic level.
Despite practical importance and broad interest to the HM ferromagnetism in CrO$_2$,
the problem was far from being fully understood. There are several reasons for it.

  First, there is no single microscopic mechanism, which is primarily responsible for the ferromagnetism of CrO$_2$.
Our analysis clearly shows that it is a joint effect of several contributions, of very different origins, and
besides conventional double exchange in the $t_{2g}$ band, there are other magnetic interactions, which are equally important
in stabilizing the ferromagnetic ground state of CrO$_2$. They include direct exchange interactions, the
interactions between $t_{2g}$ and magnetically polarized oxygen $2p$ band, as well as
higher order effects in the $(\Delta \hat{\Sigma})^{-1}$ expansion for
the magnetic energy of the $t_{2g}$ band.

  Second, the description of interatomic exchange interactions in CrO$_2$
may have many traps, because the behavior of these interactions strongly depends on the method in use, which may lead to
different conclusions.
Particularly, if one sticks to static methods, which totally neglect the effect of
correlation interactions on the magnetic properties (such as unrestricted HF approximation), the
solution of the problem may look very easy and the robust HM ferromagnetism
emerges already in the minimal model, consisting of the $t_{2g}$ bands.
However, this ``easy solution'' appears to be largely incomplete, as it becomes clear after considering the correlation interactions.
Moreover, one should be most careful with the use of additional approximations for treating the correlation interactions,
because some of these approximations may lead to unphysical results.
For instance, by using the random-phase approximation for the correlation energy and treating this problem
in the spirit of the OEP method with some \textit{static} local potential,
one can easily obtain an insulating solution, which suppresses the tendencies towards ferromagnetism.
This curious example also demonstrates the importance of dynamic correlations.

  The most reliable technique for dealing with this kind of problem is the dynamical mean-field theory.
In the present work, we have employed the new realization of this method, which is based on the
exact diagonalization solution of the quantum impurity problem, performed `on-the-fly'.
The use of this numerically advanced algorithm
enabled us not only no to solve the standard DMFT equations, but also to study in many details the behavior of
interatomic exchange interactions in the frameworks of this method. Our study provides an important insight into the
origin of HM ferromagnetism in CrO$_2$. It clearly shows that, besides conventional processes, related to the change
of the kinetic energy of electrons in the $t_{2g}$ band,
the realistic microscopic model for CrO$_2$ should also include the
direct exchange interactions and the magnetic polarization of the oxygen $2p$ band.
Finally, we have proposed how the latter two contributions can be evaluated using results of electronic structure calculations in the
local-spin-density approximation. Thus, our work provides the firm microscopic basis for understanding the magnetic
properties of CrO$_2$ -- the canonical and technologically important half-metallic ferromagnet.

\textit{Acknowledgements}.
This work is partly supported by the grant of Russian Science Foundation (project No. 14-12-00306).


\begin{thebibliography}{99}
\bibitem{note1}
All model parameters are available upon request.
\end{thebibliography}
\end{document}